\documentclass[conference]{IEEEtran}
\usepackage{cite}
\usepackage{xurl}
\usepackage{hyperref}
\usepackage{graphicx}
\usepackage{booktabs}
\usepackage{amsmath}
\usepackage{amssymb}
\usepackage{listings}
\usepackage{xcolor}
\usepackage{doi}
\usepackage{subcaption}
\usepackage{tabularx}
\usepackage{pifont} 

\begin{document}

\pagestyle{plain}

\title{Emerging Personas and Tools for\\ Hybrid Quantum-HPC Systems}
%\title{Personas, Use Cases, Tools, and Emerging Roles for\\ Hybrid Quantum-HPC Systems}

\author{
\IEEEauthorblockN{
Eun-Kyung Lee\IEEEauthorrefmark{1}\IEEEauthorrefmark{3}
Jessie Yu\IEEEauthorrefmark{1}
Claudio Carvalho\IEEEauthorrefmark{1}
Yoonho Park\IEEEauthorrefmark{1}
Marcelo Amaral\IEEEauthorrefmark{1}
Tim Osborne\IEEEauthorrefmark{2}
Woong Shin\IEEEauthorrefmark{2}
}

\IEEEauthorblockA{
\IEEEauthorrefmark{1}IBM
\IEEEauthorrefmark{2}Oak Ridge National Laboratory \\
\IEEEauthorrefmark{3}Corresponding author: eunkyung.lee@us.ibm.com
}
}

\maketitle

\thispagestyle{empty}

\begin{abstract}
Quantum computing is entering an era where quantum processing units (QPUs) with tens to hundreds of qubits are increasingly integrated with high-performance computing (HPC) systems to support hybrid quantum-classical workloads. This paradigm, often referred to as Quantum-Centric SuperComputing (QCSC), introduces new operational, software, and infrastructure challenges. In this paper, we identify and characterize six distinct personas involved in QCSC environments. Particular attention is given to emerging roles that arise at the interface between quantum and HPC systems. We analyze the responsibilities and tools associated with each persona and examine their differing requirements for operational data. Consistent data representations and analytics frameworks will be required to support diverse users across the quantum-HPC ecosystem. By defining personas and their data needs, we provide a foundation for future discussions in this critical area.
%Quantum computing is transitioning from isolated laboratory devices toward integrated components of high-performance computing (HPC) infrastructure, a paradigm that IBM has termed Quantum-Centric SuperComputing (QCSC). In these systems, quantum systems act as accelerators tightly coupled to classical supercomputers rather than standalone machines. As these systems mature, it is important to understand the different personas and tools required to fully operationalize them. In this paper we characterize six personas and the tools they use to better understand their data requirements. We emphasize three unique personas that have received comparatively little attention: the QCSC Operator, the QCSC System Engineer and the QCSC HPC Software Developer.
\end{abstract}


\section{Introduction}

Quantum computing is in pre-fault tolerant era, in which processors of hundreds of qubits are available~\cite{earlyFTQC2024}.
%A recurring lesson of this era is that quantum processors are rarely useful in isolation.
Almost every practically interesting workload from molecular ground-state estimation to combinatorial optimization interleaves quantum circuit execution with substantial classical computation for problem setup, parameter optimization, error mitigation, and results analysis. This observation motivates the tight integration of quantum processing units (QPUs) with classical HPC systems, a vision that IBM and others describe as Quantum-Centric SuperComputing (QCSC) or hybrid quantum-HPC (QHPC) systems~\cite{seelam2026qcsc}.

\begin{figure}[t]
\vspace*{-0.5pt}
\centering
\includegraphics[width=1\columnwidth]{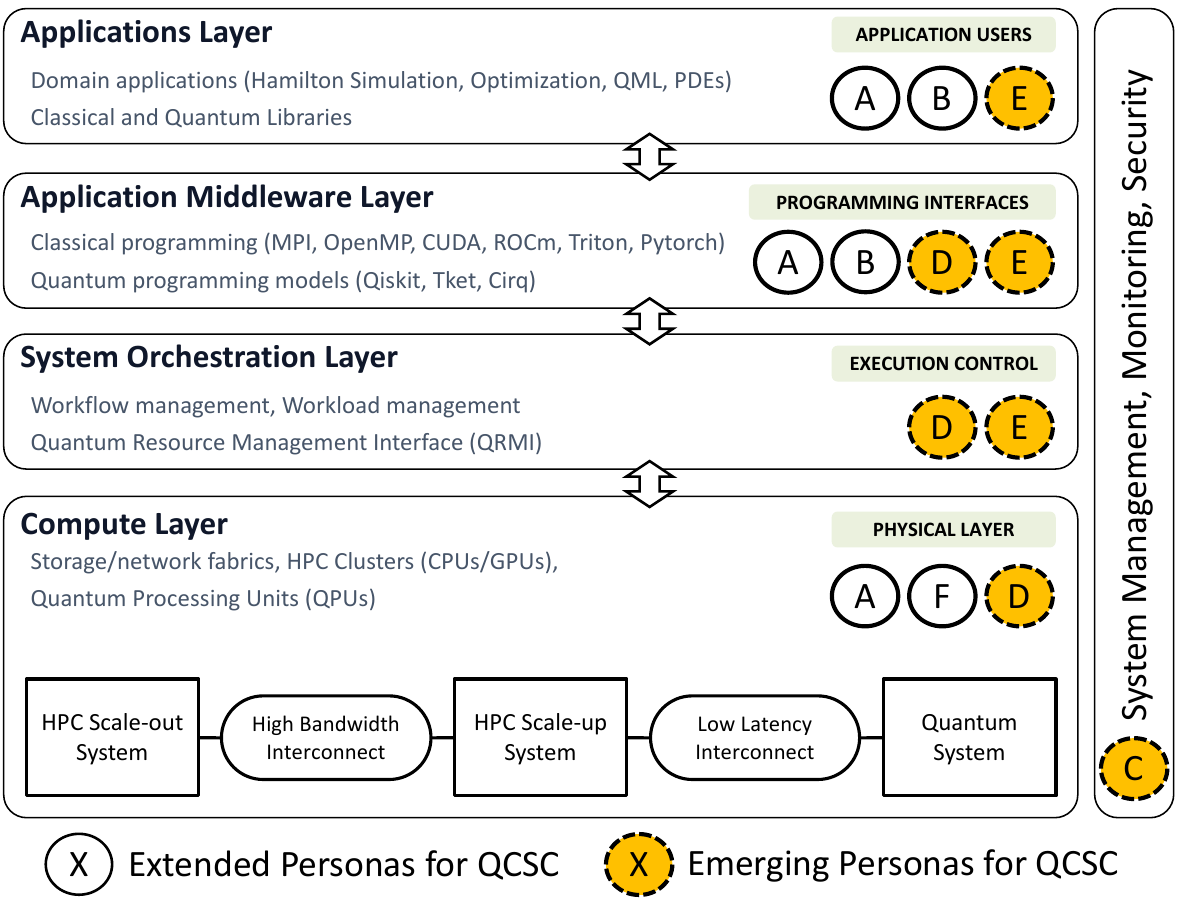}
% \caption{Overview of six quantum user personas: A.~Quantum Kernel and Library Developers, B.~Domain Scientists, C.~QCSC Operators, D.~QCSC System Engineers, E.~QCSC HPC Software Developers, and F.~Quantum Physicists. The three emerging QCSC roles are highlighted in yellow.}
\caption{Overview of six quantum user personas:  A. Quantum Kernel and Library Developers, B. Domain Scientists, C. QCSC Operators, D. QCSC System Engineers, E. QCSC HPC Software Developers, and F. Quantum Physicists. The three emerging QCSC roles are highlighted in yellow.}
\vspace{-8pt}
\label{fig:personas}
\end{figure}

% As quantum hardware matures, HPC centers such as J\"ulich, LRZ, RIKEN, and NERSC are deploying QPUs alongside flagship supercomputers~\cite{riken2026roquo,lrz2026,nersc2026}. This integration creates a demand for expertise that covers both the traditional HPC community and the quantum community. In particular, a few roles emerge at the boundary between the two different systems: the QCSC Operator, the QCSC System Engineer, and the QCSC HPC Software Developer. The QCSC Operator coordinates both quantum and HPC infrastructure. The QCSC System Engineer builds the interfaces and co-designs the hardware and software stack that couple the quantum and HPC infrastructure. The QCSC HPC Software Developer designs and implements hybrid software stacks to support other personas defined in Figure~\ref{fig:personas}.
As quantum hardware matures, HPC centers such as RIKEN, RPI, LRZ, and NERSC are deploying QPUs alongside supercomputers~\cite{riken2026roquo,rpi2025,lrz2026,nersc2026}. This integration creates a demand for expertise that spans both the traditional HPC and quantum communities. In particular, several roles emerge at the boundary between the two systems: the QCSC Operator, the QCSC System Engineer, and the QCSC HPC Software Developer. The QCSC Operator coordinates quantum and HPC infrastructure. The QCSC System Engineer builds the interfaces and co-designs the hardware and software stack that couple the two infrastructures. The QCSC HPC Software Developer designs and implements hybrid software stacks to support the other personas defined in Figure~\ref{fig:personas}. 

More broadly, QCSC systems bring together several areas, including quantum hardware, quantum software, HPC systems, and scientific applications. As a result, the people working with these systems can have different backgrounds, responsibilities, and technical needs. For example, Domain Scientists focus on solving a scientific problem, while Quantum Physicists focus on device calibration, errors, and performance. Thus, using a single term such as ``quantum user'' does not capture the range of roles. Defining clear personas helps describe who interacts with the system, what tasks they perform, and what tools and interfaces they need. This is important for understanding how QCSC systems should be designed and how their tools can better support the communities involved.

% The three emerging roles are essential to the realization of QCSC. The QCSC Operator must minimize downtime and maximize utilization across the quantum and HPC systems. The QCSC System Engineers provide the foundational interface between the two systems. They design and implement the standardized interfaces, middleware, and control frameworks that enable seamless system integration to efficiently control two systems by managing hybrid job submission and data movement between systems. QCSC HPC Software Developers bridge the software gap by translating high-level quantum algorithms into practical workflows that efficiently connect quantum and classical computation, accounting for the unique constraints of each system.

These personas also have different requirements for operational data and for the tools used to collect, analyze, and consume it. Operational Data Analytics are integral across HPC sites~\cite{ott2020global}, but data at these sites has developed organically and does not yet have a common standard describing it. This challenge becomes more important in QCSC, where data must support users and tools across both quantum and HPC systems.

% Operational Data Analytics are integral across HPC sites ~\cite{ott2020global}. Data at these HPC sites came about organically and does not yet have a standard describing it. In this paper, we identify and describe six distinct QCSC personas and tools used by the personas in order to illuminate what data is required. As we move to QCSC, data will need a consistent format throughout its lifecycle to allow different personas to utilize the data for their own needs. We hope to start the discussion in developing consistent data formats based on the personas and tools presented in this paper.

Therefore, we identify and describe six distinct QCSC personas and the tools they use to better understand their data requirements. As we move toward QCSC, this data will need a consistent format throughout its lifecycle so that different personas can use it for their own needs. We hope to start the discussion toward developing consistent data formats based on the personas and tools presented in this paper.
\section{Quantum User Personas}
\label{sec:personas}

Although no single peer-reviewed persona taxonomy for QCSC personas or roles exists, the breadth of skills required for a quantum workforce are identified in~\cite{Qskills2024}. We discuss six user personas based on motivations, goals, and tooling requirements. Our discussion is focused on the user personas emerging as quantum systems are being integrated in HPC data centers. Note that we do not discuss personas involved in the design and building of quantum systems as these personas are outside of the scope of the paper. Figure~\ref{fig:personas} illustrates the six personas and their relationship to the three QCSC-specific personas. Table~\ref{tab:tools} lists the tools used by each persona for easy comparison.

\subsection{Quantum Kernel and Library Developers}

%Quantum Kernel and Library Developers employ a rich ecosystem of tools for circuit design, optimization, and algorithm development. Key frameworks include: (1) Circuit Compilation Platforms (Qiskit~\cite{qiskit2024}, Cirq~\cite{cirq2024}, and PyQuil~\cite{smith2016practical}) for abstract circuit design, optimization, and compilation to hardware-specific gate sets; (2) Error Mitigation Techniques including probabilistic error cancellation~\cite{van_den_Berg_2023}, zero-noise extrapolation, and sparse Pauli-Lindblad models for extending quantum computer reach on pre-fault tolerant devices; (3) Variational Quantum Algorithm (VQA) Design Tools~\cite{cerezo2021variational2} including ansatz optimization frameworks for hardware-efficient circuits; and (4) Benchmarking and Validation Frameworks~\cite{benchmarking2025} for assessing algorithm performance, circuit fidelity, and system characterization.

Quantum Kernel and Library Developers are experts in quantum computing. They research circuit methodologies and algorithms. Their goal is to optimize circuits for performance, fidelity, and resource efficiency - for example, designing hardware-efficient ans\"atze for variational algorithms~\cite{kandala2017hardware} or error-mitigation schemes that extend the computational reach of near term devices~\cite{kandala2019errormit}.
Quantum Kernel and Library Developers view QCSC convergence from an algorithmic efficiency standpoint, serving as the architects of performance in hybrid quantum-classical workflows. Their perspective emphasizes: 
\begin{itemize}
    \item Circuit Optimization: Design large-scale circuits that execute efficiently on noisy quantum hardware to obtain the highest-quality results.
    \item Error Mitigation at Scale: Develop mitigation strategies that remain practical when hybrid workloads demand many repeated quantum executions integrated into classical HPC jobs.
    \item Hybrid Algorithm Optimization: Develop or improve hybrid quantum-classical algorithms, ensuring hybrid workflows achieve better performance than purely classical or purely quantum solutions.
    \item Fault-Tolerant Building Blocks: Develop and benchmark error-correction primitives (encoding, syndrome extraction, decoding) that library consumers can compose, working toward systems that correct errors faster than they accumulate. 
\end{itemize}
Note that Fault-Tolerant Building Blocks is at the Compute Layer in Figure~\ref{fig:personas} while Circuit Optimization, Error Mitigation, and Hybrid Algorithm Optimization are at the Application Layer and Application Middleware Layer.

\subsection{Domain Scientists}

%We distinguish two sub-personas: academic researchers and industry researchers. 
%Academic Researchers are discovery-oriented and pursue longer term problems such as quantum simulation for life sciences and materials, where the aim is scientific insight rather than immediate profit. They employ domain-specific simulation frameworks such as OpenFermion for electronic structure problems~\cite{cao2019quantum,reiher2020elucidating}, molecular dynamics libraries, and materials science simulation tools~\cite{mcarthur2023quantum}. Industry researchers are more focused on practical areas such as optimization, logistics, finance, and security, where measurable business value drives adoption. They employ quantum optimization toolkits including QAOA and VQE for combinatorial problems~\cite{zhou2023quantum}, warm-starting techniques for improved convergence~\cite{egger2023quantum}, and machine learning frameworks for feature extraction and pattern recognition~\cite{liu2024quantum}.

% They leverage cloud quantum platforms (IBM Quantum, AWS Braket, Azure Quantum~\cite{ibmquantum2024,awsbraket2024,azurequantum2024}) to access diverse quantum hardware modalities, and utilize parameterized circuit frameworks like TensorFlow Quantum~\cite{tensorflow2024quantum} for variational simulations.
% Industry teams utilize application-focused frameworks that abstract quantum complexity~\cite{teplitskiy2023quantum}, enable rapid prototyping, and integrate with existing enterprise systems and HPC infrastructure~\cite{ferracin2024quantum}.

Domain Scientists are experts in non-quantum domains, such as chemistry or machine learning. They are not quantum experts, but they do know how to apply quantum algorithms to solve application-specific problems; they care about results, not circuit internals~\cite{merz2026crossing12000atombarrierheterogeneous}.
Domain Scientists view QCSC convergence from an application-first standpoint, prioritizing end-to-end solution quality and time-to-solution over quantum algorithmic elegance. Their perspective emphasizes: 
\begin{itemize}
    \item Business/Scientific Value Maximization: Design hybrid workflows by identifying which sub-problems can be solved more efficiently with quantum algorithms, while the rest is processed by classical HPC, yielding better wall-clock performance than either alone.
    \item Practical Hardware Flexibility: Design application logic that adapts to available quantum hardware characteristics (qubit count, connectivity, gate fidelity, execution speed) rather than requiring specific idealized platforms. This enables portability across different QCSC systems.
    \item Cost-Performance Trade-offs: Balance quantum execution time against classical compute cost to achieve an acceptable time-to-solution within a resource budget.
\end{itemize}

%(4) \textbf{Incremental Problem Scaling}---starting with toy-scale problems to validate hybrid workflows on available hardware, then progressively scaling to production instances as quantum hardware improves, ensuring continuous value delivery throughout the quantum technology roadmap.

\subsection{QCSC Operators}

%QCSC Operators employ a comprehensive toolkit for managing QCSC environments. Key resources include: (1) Monitoring and Observability Frameworks~\cite{prometheus2024,grafana2024} for metrics collection and visualization, augmented with quantum specific health monitoring systems that track qubit coherence, gate fidelity, and device availability; (2) Resource Scheduling and Partitioning (e.g. SLURM) configured to manage heterogeneous quantum-classical workloads with quantum-aware partition policies and reservation systems; (3) Access Control and Authentication (industry-standard frameworks including OpenID Connect, Keycloak, and HashiCorp Vault~\cite{oidc2024,keycloak2024,vault2024}) for federated identity management and secure credential handling across quantum and HPC resources; (4) Fault Detection and Recovery Mechanisms (automated systems for detecting quantum device failures, isolating faulty resources, and executing recovery procedures); (5) Configuration Management and Infrastructure-as-Code (Terraform, Ansible, and containerization platforms (Docker)~\cite{terraform2024,ansible2024,docker2024}) for reproducible system deployment and version-controlled infrastructure specifications; and (6) Capacity Planning Tools (frameworks for analyzing workload patterns and optimizing resource allocation across quantum and classical domains).

QCSC Operators coordinate both quantum and HPC infrastructure. Their focus is minimizing downtime and maximizing utilization, configuring scheduler partitions, defining access and reservation policies, mapping credentials to devices, and monitoring the health of heterogeneous resources~\cite{ott2020global}.
QCSC Operators view QCSC systems from a security, operational reliability, and multi-user efficiency standpoint. Their perspective emphasizes:
\begin{itemize}
    \item Security: Reconciling the quantum and HPC infrastructure security models. Ensuring credentials are granted and stored appropriately. This perspective is discussed in more detail in~\cite{badts2026ExaminingQRMIunifiedInterface} with respect to workload managers for QCSC systems.
    \item Unified Monitoring and Alerting: Providing end-to-end visibility into QCSC workload performance, quantum device health, and classical resource utilization through integrated dashboards that alert operators to anomalies in either domain.
    \item Optimized Resource Utilization: Balancing quantum and classical resource allocation to maximize system throughput while preventing bottlenecks, ensuring expensive quantum resources are efficiently utilized and classical compute is optimally provisioned.
    \item Multi-User Access and Fairness: Managing fair resource sharing among competing users and applications through scheduling policies that prevent quantum-resource starvation and ensure predictable access and performance isolation.
    \item Resilience and Availability: Designing systems that gracefully handle quantum device failures (e.g., qubit decoherence, gate errors) and classical resource failures, maintaining service continuity through redundancy, failover mechanisms, and rapid recovery procedures essential for mission-critical QCSC deployments.
\end{itemize}

\subsection{QCSC System Engineers}

%QCSC System Engineers employ a sophisticated toolkit for QCSC system integration spanning hardware abstraction, software stack, and standardization. Key resources include: (1) Hardware Abstraction and Device Drivers (OpenQASM 3~\cite{openqasm2024} and Quantum Intermediate Representation (QIR)~\cite{qir2024}) for hardware-agnostic quantum program representation, enabling portability across heterogeneous quantum platforms; (2) Software Stack Co-design Tools (compiler infrastructure like LLVM for quantum, TVM for optimization~\cite{tvm2024quantum}, and high-level languages like Silq~\cite{silq2024}) for type-safe quantum program development and automated compilation to device-specific gates; (3) Middleware Frameworks for transparent communication between quantum and classical components, managing data serialization, error propagation, and resource coordination; (4) Network Infrastructure and Interconnects (gigabit-speed quantum-classical links and standardized interfaces) for integrating quantum processors into supercomputer architectures; (5) Quantum Simulation and Emulation Tools (platforms like qsim~\cite{qsim2024} and ProjectQ~\cite{quantumsim2024}) for full-system simulation and validation before deployment; and (6) Hardware-Software Co-design Frameworks (design methodologies and tools) for jointly optimizing quantum device control, classical runtime systems, and hybrid execution models.

QCSC System Engineers build the interfaces that bridge quantum and HPC systems. They co-design the hardware and software stack (middleware, firmware, and network paths) so that a QPU appears as a managed, schedulable resource~\cite{resourceQuantumInterconnectNetworks2026,badts2026ExaminingQRMIunifiedInterface}. This persona is more focused on the hardware and software integration challenges than the QCSC HPC Software Developer who is focused on hybrid workflow design and optimization.
QCSC System Engineers approach QCSC convergence from a \textit{system architecture and standardization} standpoint, focusing on seamless quantum-classical integration. Their perspective emphasizes: 
\begin{itemize}
    \item Hardware Abstraction and Portability: Design standardized interfaces that decouple quantum algorithms from specific hardware, enabling QCSC systems to support multiple quantum modalities such as superconducting, trapped ion, photonic without middleware changes.
    \item Unified Software Stack: Co-design compiler, runtime, and middleware layers that treat quantum and classical execution as a unified computational model rather than separate domains, minimizing latency and synchronization overhead.
    \item Interoperability and Standardization: Drive adoption of cross-platform standards that enable independent development of quantum and classical components while ensuring seamless integration at the QCSC system level.
    \item Scalability and Heterogeneity: Design systems that gracefully scale from single quantum processors to distributed QPU networks across supercomputer infrastructure, supporting diverse hardware characteristics and performance profiles while maintaining a coherent programming and execution model.
\end{itemize}

\subsection{QCSC HPC Software Developers}

%QCSC HPC Software Developers rely on a specialized ecosystem of integration tools and frameworks. Key resources include: (1) Resource Management Interface: Quantum Resource Management Interface (QRMI)~\cite{qrmi2025} and Quantum Device Management Interface (QDMI)~\cite{qdmi2024} for unified resource provisioning and scheduling across heterogeneous quantum and classical systems; (2) Workflow Orchestration Frameworks: Apache Airflow, Nextflow, and similar tools for coordinating multi-step hybrid jobs~\cite{workflows2024}; (3) Hybrid Programming Models and reference architectures for quantum-centric supercomputers~\cite{seelam2026qcsc}; (4) Data Movement Libraries optimizing quantum-classical communication and minimizing latency overhead; (5) Performance Profiling and Optimization Tools for analyzing hybrid workload characteristics and identifying bottlenecks; and (6) Quantum-Ready Primitives: high-level abstractions (e.g., Hamiltonian simulation blocks, optimization routines) that enable application developers to compose hybrid workflows without detailed quantum knowledge~\cite{delgado2025primitives}.

QCSC HPC Software Developers are experts in classical software development. They ensure quantum-classical hybrid workloads can run efficiently in a heterogeneous compute environment. They express workflows in terms of classical pre-processing, quantum execution, and classical post-processing, requiring deep understanding of available compute resources, resource reservation, and quantum-classical data movement~\cite{seelam2026qcsc,humble2021qchpc}.
QCSC HPC Software Developers view QCSC convergence from a workflow integration standpoint, serving as the primary integrators bridging quantum and HPC software ecosystems. Their perspective emphasizes: 
\begin{itemize}
    \item Latency Minimization: Reduce round-trip communication overhead between classical and quantum resources through asynchronous execution, batching, and in-situ processing.
    \item Resource Efficiency: Ensure hybrid jobs make optimal use of both quantum and classical compute capacity without idle periods or resource bottlenecks, critical for converged supercomputing environments with shared multi-user resources.
    \item Scheduling and Orchestration: Design workload submission patterns that integrate with HPC job schedulers, respect resource constraints, and achieve acceptable queue times for hybrid applications.
    \item Portability: Abstract quantum hardware heterogeneity and allow hybrid applications to run across different QCSC systems and quantum device modalities without code modification.
\end{itemize}

\subsection{Quantum Physicists}

%This diagnostic work is inherently modality-specific, since the dominant error mechanisms differ markedly between superconducting, trapped-ion, and photonic platforms.
%superconducting circuits~\cite{kjaergaard2020superconducting} and trapped ions~\cite{bruzewicz2019trappedion}.

%Quantum Physicists rely on a targeted set of tools for probing, characterizing, and calibrating existing hardware. Key frameworks include: (1) Qiskit Pulse and Qiskit Dynamics~\cite{qiskit2023dynamics} for constructing custom pulse sequences and simulating time-dependent Hamiltonian evolution directly on device hardware to expose gate errors and decoherence sources; (2) QuTiP~\cite{johansson2013qutip}, a Python framework for open-quantum-systems simulation that models decoherence mechanisms and identifies error channels on physical devices; (3) Quantum Characterization, Verification, and Validation (QCVV) techniques including randomized benchmarking~\cite{knill2008benchmarking} and gate set tomography for systematic error diagnosis and identification of corrective calibration steps; and (4) Hardware control frameworks such as ARTIQ and QICK for executing diagnostic pulse sequences with nanosecond-level timing precision and real-time feedback on deployed quantum devices, enabling iterative identification and mitigation of gate errors across diverse quantum platforms.

Quantum Physicists have deep understanding of device-level physics. They use pulse-level control to diagnose device errors and identify fixes on deployed hardware~\cite{quantumInformationHardwareChallenges2025}. Rather than building hardware, their focus is probing/tuning what goes wrong: characterizing decoherence sources, isolating error types, and determining corrective pulse parameters to restore fidelity.
Quantum Physicists view QCSC convergence from a device quality standpoint, acting as the feedback loop that keeps deployed QPUs performant within the broader HPC environment. Their perspective emphasizes:
\begin{itemize}
    \item Error Feedback from Hybrid Workloads: Correlating degraded hybrid job outcomes with device-level error signatures. For example, how gate errors propagate through hybrid QCSC jobs and how feedback from failed QCSC jobs informs device tuning. 
    \item Infrastructure Integration Challenges: Characterize how devices perform under HPC operating conditions (shared resource contention, queuing delays, thermal management in data centers, etc.) rather than just under laboratory conditions.
    \item Scalability Concerns: Moving from small-scale devices to ones integrated into supercomputer systems requires innovation in calibration efficiency and in error identification and handling. 
\end{itemize}

\begin{table*}[t]
  \begin{tabularx}{\textwidth}{>{\raggedright\arraybackslash}p{6cm}X}
    \toprule
    Persona & Tools and Frameworks \\
    \midrule
    A. Quantum Kernel and Library Developers & Circuit compilation and execution platforms for abstract circuit design, optimization, and compilation to hardware-specific gate sets; error mitigation frameworks for extending quantum computer reach on pre-fault tolerant devices; benchmarking and validation frameworks for assessing algorithm performance, circuit fidelity, and system characterization \\
    \midrule
    B. Domain Scientists & Domain-specific simulation frameworks such as OpenFermion~\cite{openfermion2020} for electronic structure problems; molecular dynamics libraries; materials science simulation tools; quantum optimization toolkits including QAOA~\cite{farhi2014qaoa} and VQE~\cite{cerezo2021variational2} for combinatorial problems; machine learning frameworks for feature extraction and pattern recognition \\
    \midrule
    C. QCSC Operator & Monitoring and observability frameworks for metrics collection and visualization; workload managers configured to manage heterogeneous quantum-classical workloads; access control and authentication frameworks for federated identity management and secure credential handling across quantum and HPC resources; fault detection and recovery mechanism frameworks; configuration management and Infrastructure-as-Code tools for reproducible system deployment and version-controlled infrastructure specifications; capacity planning tools for analyzing workload patterns and optimizing resource allocation \\
    \midrule
    D. QCSC System Engineers & Resource management interface such as QRMI~\cite{bacher2025quantumresourcesresourcemanagement} and QDMI~\cite{qdmi2024}; hardware abstraction frameworks for hardware-agnostic quantum program representation; software stack co-design tools such as LLVM for quantum program development and automated compilation to device-specific gates; middleware frameworks for communication between quantum and classical components; network infrastructure for integrating quantum processors into supercomputer architectures; quantum simulation for full-system simulation and validation before deployment \\
    \midrule
    E. QCSC HPC Software Developers & Resource management interface such as QRMI and QDMI; workflow tools such as Nextflow~\cite{nextflow2017} and Snakemake~\cite{snakemake2021} for coordinating multi-step hybrid jobs; hybrid programming tools; data movement libraries optimizing quantum-classical communication and minimizing latency overhead; performance profiling and optimization tools for analyzing hybrid workload characteristics and identifying bottlenecks~\cite{nation2026,walkup2026}; quantum primitives that enable application developers to compose hybrid workflows without detailed quantum knowledge \\
    \midrule
    F. Quantum Physicists & Probing tools for constructing custom pulse sequences; quantum characterization, verification, and validation (QCVV) toolkits; hardware control frameworks for executing probing tools \\
    \bottomrule
  \end{tabularx}
  \caption{Tools and frameworks used by each persona.}
  \label{tab:tools}
\end{table*}

%\subsection{HPC vs. QCSC User Personas}
%Traditional HPC personas have well-established for many decades, but QCSC roles differ in important ways, summarized in Table~\ref{tab:roles}. Classical HPC users reason about homogeneous or CPU-GPU-accelerated nodes with mature, deterministic scheduling. QCSC users must additionally reason about scarce, noisy, non-deterministic QPUs of different modalities, each with distinct error characteristics, calibration cycles, and vendor APIs.

%\begin{table}[t]
%\centering
%\caption{Traditional HPC roles versus QCSC roles.}
%\label{tab:roles}
%\begin{tabular}{@{}p{2.3cm}p{2.6cm}p{2.6cm}@{}}
%\toprule
%\textbf{Dimension} & \textbf{HPC role} & \textbf{QCSC role} \\
%\midrule
%Software Developer & Parallelizes code (MPI/OpenMP) on CPU/GPU & Partitions workload into classical/quantum stages; manages QPU calls via QRMI \\
%Admin & Schedules jobs on homogeneous nodes & Schedules scarce, noisy QPUs ; maps credentials and reservations \\
%Engineer & Integrates interconnects, storage, accelerators & Builds QPU middleware/plugins; co-designs quantum-classical interface \\
%Resource model & Deterministic, abundant cores & Non-deterministic, scarce, modality-specific QPUs \\
%\bottomrule
%\end{tabular}
%\end{table}

%\input{sections/sec3.modalities}
%\input{sections/sec4.hybrid_applications}
\section{Discussion}
\label{sec:discussion}

% QCSC data centers are incredibly complex, and their users have vastly different goals, technical skills, and workflows. On the other hand, QCSC users range from personas focused on domain-specific problems to those focused on the Quantum hardware. Failing to understand personas can lead to QCSC data centers that are too difficult for one user persona or too restrictive others.

The personas and tools described in Section~\ref{sec:personas} focused on a subset of QCSC users that are critical for QCSC adoption. Of course, data from QCSC systems will be used by a broader audience. The set of personas described in this paper should be expanded both ``vertically'' and ``horizontally.'' By ``vertically'' we mean expanding the set according to how the QCSC systems are funded, procured, operated, and used. This could bring in Sponsor and Program Manager personas. By ``horizontally'' we mean subdividing personas such as QCSC Operator into more specific personas such as System Administrator, User Support, and Accounting. Of course, expanding set of personas ``vertically'' and ``horizontally'' requires a discussion with a wider group of experts.

\begin{table}[t]
\centering
\caption{Data categories and relevant user personas.}
\label{tab:discussion}
\begin{tabular}{|p{3.5cm}|c|c|c|c|c|c|}
% \hline
%  & \multicolumn{5}{c|}{Some text} & \multicolumn{4}{c|}{Some text} \\
\hline & \textbf{A} & \textbf{B} & \textbf{C} & \textbf{D} & \textbf{E} & \textbf{F} \\
\hline
\multicolumn{7}{|l|}{\textbf{Facility Data}} \\
\hline
\quad Cooling data          &  &  &  & \checkmark  &  & \checkmark \\
\quad Power data            &  &  &  & \checkmark & \checkmark & \checkmark \\
\quad Cost data             &  &  &  & \checkmark & \checkmark &  \\
\hline
\multicolumn{7}{|l|}{\textbf{System Hardware Telemetry Data}} \\
\hline
\quad Resource data         &  &  & \checkmark & \checkmark & \checkmark &  \\
\quad Capacity data         &  &  & \checkmark & \checkmark & \checkmark &  \\
\quad Q-Control data        & \checkmark &  &  &  &  & \checkmark \\
\quad Q-Fidelity data       & \checkmark & \checkmark &  &  &  & \checkmark \\
\quad Q-Calibration data    & \checkmark &  &  &  &  & \checkmark \\
\quad Maintenance data      & \checkmark & \checkmark & \checkmark & \checkmark & \checkmark & \checkmark \\
\hline
\multicolumn{7}{|l|}{\textbf{System Software Scheduler Data}} \\
\hline
\quad Job data              &  &  &  &  & \checkmark &  \\
\quad Queuing data          &  &  &  &  & \checkmark &  \\
\quad Scheduling data       &  &  & \checkmark & \checkmark & \checkmark &  \\
\quad Various job timing data   &  &  & \checkmark & \checkmark & \checkmark &  \\
\hline
\multicolumn{7}{|l|}{\textbf{Workflow Data}} \\
\hline
\quad Application-specific data  &  & \checkmark &  &  &  &  \\
\quad Multi-system I/O profiles &  & \checkmark &  & \checkmark & \checkmark &  \\
\hline
\end{tabular}
\end{table}

Table~\ref{tab:discussion} summarizes the data categories that each of the six user personas may need to access or manage. The table is organized into four main data categories, each with specific types of data relevant to the different personas. The checkmarks indicate which personas are likely to interact with or require access to each type of data. \textit{Facility Data} includes cooling, power, and cost data at various levels (datacenter, system, rack, node). \textit{System Hardware Telemetry Data} includes resource data (CPU/QPU/GPU utilization, storage utilization, network utilization), capacity data (cloud resources), quantum control and fidelity (error) data, quantum calibration data (coherence time, quantum fidelity, readout error), and maintenance data (maintenance history, scope of impact). \textit{System Software Scheduler Data} includes job data (job and associated resource requirements), job queuing data, scheduling/allocation data, and various job timing data (start/end time, pending/wait time). \textit{Workflow Data} include application-specific data (application throughput, checkpoints, multi-system throughput). The primary purpose of those data collection in QCSC data center is to maximize computational throughput while minimizing energy/operational costs and preventing hardware failure. 

The three emerging QCSC personas (QCSC Operator, QCSC System Engineer, and QCSC HPC Software Developer) become more critical for several reasons. First, a single center may host different Quantum system modalities (superconducting, trapped-ion, and neutral-atom), each with its own API, error profile, and calibration schedule, multiplying the integration and administration burden. Second, correctness of hybrid results depends on well-designed classical post-processing and error mitigation. Third, portability becomes essential and vendor-agnostic interfaces such as QRMI~\cite{qrmi2025} protect workflows from hardware changes. Fourth, utilization matters more as QPUs remain scarce and expensive. These integration challenges fall in the domain of the three emerging QCSC personas.

%As HPC centers deploy an increasing number and variety of QPUs, the three emerging QCSC personas become more critical for several reasons. First, a single center may host different modality Quantum system (superconducting, trapped-ion, and neutral-atom), each with its own API, error profile, and calibration schedule, multiplying the integration and administration burden. Second, correctness of hybrid results depends on well-designed classical post-processing and error mitigation. Third, portability becomes essential and vendor-agnostic interfaces like QRMI~\cite{qrmi2025} protect workflows from hardware changes. Each of the reasons falls within the domain of the QCSC Operator, the QCSC System Engineer, or the QCSC HPC Software Developer. Furthermore, utilization matters more as QPUs remain scarce and expensive. Combined efforts of those three personas are needed to schedule classical and QPUs efficiently and avoid idle time. 

%These personas are especially critical for ensuring that hybrid workflows complete within acceptable time windows, optimizing resource utilization across quantum and classical systems, maintaining reliability in mixed-hardware environments, and translating scientific requirements into operational policies. The emergence of reference architectures and standardized interfaces for QCSC~\cite{alexeev2021qcsystems,humble2021qchpc,delgado2025primitives} signals that these personas are becoming a recognized part of the HPC operations landscape rather than an ad hoc afterthought.

\section{Conclusion}
This paper clarified the use cases of quantum systems through the lens of six user personas and emphasized three roles that sit at the quantum-HPC boundary: QCSC Operators, QCSC System Engineers, and QCSC HPC Software Developers. As QCSC systems mature, it is essential to understand the distinct personas involved and the tools they require in order to fully operationalize QCSC systems. Efficient operationalization depends on data. A better understanding of the personas and tools will help the community to develop consistent data formats.

%Finally, we contrasted traditional HPC and QCSC roles and argued that, as HPC-quantum systems proliferate across modalities and centers, these specialized roles will be essential to realizing the promise of quantum-accelerated computing.

\section*{Acknowledgments}
The authors used AI-assisted tools, specifically, Claude and Copilot, during the preparation of this work to support tasks such as grammar checking, literature summarization, and text drafting. All AI-generated content was reviewed, edited, and verified by the authors. The authors take full responsibility for the accuracy, integrity, and originality of the submitted work. AI tools were not used for analysis or the drawing of conclusions.

%\clearpage
\bibliographystyle{IEEEtran}
\bibliography{references}
\end{document}